\documentclass{aastex63}
\submitjournal{ApJL}
\shorttitle{GRB~220426A: a burst similar to GRB~090902B}
\shortauthors{Deng et al.}
\newcommand{\MyFigA}{\ref{MyFigA}}
\newcommand{\MyFigB}{\ref{MyFigB}}
\newcommand{\MyFigC}{\ref{MyFigC}}
\newcommand{\MyFigD}{\ref{MyFigD}}
\newcommand{\MyFigE}{\ref{MyFigE}}
\newcommand{\MyFigF}{\ref{MyFigF}}

\newcommand{\MyTabA}{\ref{MyTabA}}

\begin{document}
\title{\emph{Fermi} Observations of GRB~220426A: a burst similar to GRB~090902B}
\correspondingauthor{Da-Bin Lin}
\email{lindabin@gxu.edu.cn}
\author{Li-Tao Deng}
\affil{Laboratory for Relativistic Astrophysics, Department of Physics, Guangxi University, Nanning 530004, China}
\author{Da-Bin Lin}
\affil{Laboratory for Relativistic Astrophysics, Department of Physics, Guangxi University, Nanning 530004, China}
\author{Li Zhou}
\affil{Laboratory for Relativistic Astrophysics, Department of Physics, Guangxi University, Nanning 530004, China}
\author{Kai Wang}
\affil{Laboratory for Relativistic Astrophysics, Department of Physics, Guangxi University, Nanning 530004, China}
\author{Xing Yang}
\affil{Laboratory for Relativistic Astrophysics, Department of Physics, Guangxi University, Nanning 530004, China}
\author{Shu-Jin Hou}
\affil{College of Physics and Electronic Engineering, Nanyang Normal University, Nanyang, Henan 473061, China; houshujingrb@163.com}
\affil{Laboratory for Relativistic Astrophysics, Department of Physics, Guangxi University, Nanning 530004, China}
\author{Jing Li}
\affil{Laboratory for Relativistic Astrophysics, Department of Physics, Guangxi University, Nanning 530004, China}
\author{Xiang-Gao Wang}
\affil{Laboratory for Relativistic Astrophysics, Department of Physics, Guangxi University, Nanning 530004, China}
\author{Rui-Jing Lu}
\affil{Laboratory for Relativistic Astrophysics, Department of Physics, Guangxi University, Nanning 530004, China}
\author{En-Wei Liang}
\affil{Laboratory for Relativistic Astrophysics, Department of Physics, Guangxi University, Nanning 530004, China}
\begin{abstract}
We report on a very bright, long-duration gamma-ray burst (GRB), GRB~220426A,
observed by \emph{Fermi} satellite.
GRB~220426A with total duration of $T_{90}=6$~s is composed with two main pulses and some sub-peaks.
The spectral analysis of this burst with Band function reveals that
both the time-integrated and the time-resolved spectra
are very narrow with high $\alpha \gtrsim 0.2$ and low $\beta\lesssim -3.1$.
It is strong reminiscent of GRB~090902B,
a special GRB with identification of the photospheric emission.
Then, we perform the spectral analysis of this burst
based on a non-dissipated photospheric emission,
which can be well modelled as the multicolor-blackbody with a cutoff power-law distribution of the thermal temperature.
The spectral fittings reveal that the photospheric emission can well describe the radiation spectrum of this burst.
We conclude that this burst would be a second burst in the class of GRB~090902B observed by \emph{Fermi} satellite.
We also discuss the physics of photosphere and the origin of the high-energy component in GRB~220426A .
\end{abstract}
\section{Introduction}\label{Sec:Intro}
The emission mechanism of gamma-ray bursts (GRBs) has been a puzzle since their discovery half a century ago.
The difficulty in understanding prompt emission mainly lies in
that no known theoretical models can straightforwardly interpret
all of the observational data.
According to the large statistics of the temporal and spectral properties for the prompt emission,
two main categories about the prompt emission mechanism are proposed.
One invokes the non-thermal emission mechanism,
owing to the non-thermal characteristic of the Band component observed in most of GRBs.
In this scenario, previous works have shown that the synchrotron
or synchrotron-self-Compton radiation emitted by accelerated electrons is the promising mechanism (\citealp{Tavani_M-1996-ApJ.466.768T}; \citealp{Lloyd_NM-2000-Petrosian_V-ApJ.543.722L}; \citealp{Zhang_B-2002-Meszaros_P-ApJ.581.1236Z}; \citealp{Daigne_F-2011-Bosnjak_Z-A&A.526A.110D}; \citealp{Zhang_B-2011-Yan_H-ApJ.726.90Z}; \citealp{Uhm_ZL-2014-Zhang_B-NatPh.10.351U}).
Another mechanism is a Comptonized quasi-thermal emission from the outflow photosphere
(\citealp{Thompson_C-1994-MNRAS.270.480T}; \citealp{Ghisellini_G-1999-Celotti_A-ApJ.511L.93G}; \citealp{Peer_A-2006-Meszaros_P-ApJ.642.995P}; \citealp{Thompson_C-2007-Meszaros_P-ApJ.666.1012T}; \citealp{Giannios_D_-2008-A&A.480.305G}; \citealp{Lazzati_D-2010-Begelman_MC-ApJ.725.1137L};
\citealp{Mizuta_A-2011-Nagataki_S-ApJ.732.26M}; \citealp{Lazzati_D-2013-Morsony_BJ-ApJ.765.103L}; \citealp{Ruffini_R-2013-Siutsou_IA-ApJ.772.11R}),
according to the quasi-thermal components detected in the spectrum of some GRBs (\citealp{Ryde_F-2004-ApJ.614.827R}; \citealp{Ryde_F-2005-ApJ.625L.95R}; \citealp{Ryde_F-2009-Peer_A-ApJ.702.1211R}; \citealp{Abdo_AA-2009-Ackermann_M-ApJ.706L.138A};
\citealp{Ryde_F-2010-Axelsson_M-ApJ.709L.172R}; \citealp{Zhang_BB-2011-Zhang_B-ApJ.730.141Z}; \citealp{Guiriec_S-2011-Connaughton_V-ApJ.727L.33G}; \citealp{Axelsson_M-2012-Baldini_L-ApJ.757L.31A}; \citealp{Ghirlanda_G-2013-Pescalli_A-MNRAS.432.3237G}; \citealp{Larsson_J-2015-Racusin_JL-ApJ.800L.34L}; \citealp{Guiriec_S-2013-Daigne_F-ApJ.770.32G}; \citealp{Toma_K-2011-Wu_XF-MNRAS.415.1663T}).
Although the thermal components are rarely observed,
their contributions on the GRB prompt emission could not be ignored.
Some authors have suggested that the photospheric emission is an inherent
component for the fireball model (\citealp{Meszaros_P-2000-Rees_MJ-ApJ.530.292M}; \citealp{Meszaros_P-2002-RamirezRuiz_E-ApJ.578.812M}; \citealp{Daigne_F-2002-Mochkovitch_R-MNRAS.336.1271D}; \citealp{Rees_MJ-2005-Meszaros_P-ApJ.628.847R}).
Observationally, bright photospheric emission is indeed found in several bursts,
e.g., GRB 090902B, which highlights the importance of the photospheric emission during GRB prompt phase.

GRB~090902B is a bright, long GRB, detected by the Gamma-ray Burst Monitor (GBM) and Large Area Telescope on board the \emph{Fermi} Gamma-ray Space Telescope.
The spectrum of its prompt emission is peculiar.
Some works revealed that the gamma-ray prompt emission of this burst is
dominated by the thermal emission in the energy range
from $\sim$50~keV to $\sim$10~MeV (\citealp{Abdo_AA-2009-Ackermann_M-ApJ.706L.138A}; \citealp{Ryde_F-2010-Axelsson_M-ApJ.709L.172R}; \citealp{Zhang_BB-2011-Zhang_B-ApJ.730.141Z}).
\cite{Abdo_AA-2009-Ackermann_M-ApJ.706L.138A} revealed that both the time-integrated and time-resolved
spectra of this burst can be fitted with the Band function and an additional power-law spectral component.
Compared with the Band component found in other GRBs,
however, the Band component in GRB~090902B is very narrow with $\alpha\sim -0.3$ and $\beta\sim -3.32$,
rather than general values of $\alpha\sim -1$ and $\beta\sim -2.3$ (\citealp{Preece_RD-2000-Briggs_MS-ApJS.126.19P}).
Since the shape of the Band component in this burst is too narrow for synchrotron radiation,
it is likely of a photospheric origin.
Based on the spectral model of a multi-color blackbody with a power-law distribution of temperatures,
\cite{Ryde_F-2010-Axelsson_M-ApJ.709L.172R} found that the time-resolved spectra of the prompt emission can be well fitted for the dominant component in this burst.
GRB~220426A is a bright burst with a duration $T_{90}\sim6$~s over the energy range from 50~keV to 300~keV.
In the period of [0.002, 9.856]~s after the Fermi trigger,
the fluence in the energy range of 10-1000~keV
is $(1.084\pm0.005)\times10^{-4}$~$\rm erg\cdot cm^{-2}$
and the time-integrated spectrum fitted by a Band function
reports a narrow Band component with
$E_{\rm p}= 146.3\pm 0.9$~keV, $\alpha = -0.05\pm 0.01$, and $\beta = -3.08\pm0.04$ (\citealp{Malacaria_C-2022-Meegan_C-GCN.31955.1M}).
This burst also triggered the observation of the Konus-Wind experiment (\citealp{Frederiks_D-2022-Lysenko_A-GCN.31959.1F}).
Based on the Konus-Wind observation,
the time-integrated spectral analysis of this burst also reveal
a narrow Band component with $\alpha = -0.29_{-0.07}^{+0.07}$ and $\beta =-4.00_{-0.71}^{+0.32}$ in this burst (\citealp{Frederiks_D-2022-Lysenko_A-GCN.31959.1F}).
The narrow Band component found in this burst is very similar to that found in GRB~090902B.
Then, we would like to believe that the photosphere emission is responsible for the prompt emission of GRB~220426A.

In this letter, we report on the observations and spectra analysis of GRB~220426A.
The paper is organized as follows.
In Section~\ref{Sec:spectrum_sat}, the modeling of a non-dissipated photosphere emission is presented.
Different from \cite{Ryde_F-2010-Axelsson_M-ApJ.709L.172R},
a multi-color blackbody emission model with cutoff power-law distribution of temperature is proposed.
In Section~\ref{Sec:Analysis},
the data reduction and the spectral analysis of GRB~220426A is performed.
In Section~\ref{Sec:Conclusion}, we present the conclusion and make some discussion about this burst.

\section{Spectra Modeling of a non-dissipated photospheric emission}\label{Sec:spectrum_sat}
Several GRBs are identified having a distinct thermal component by fitting with a Plank radiation spectrum.
However, the photospheric emission is generally different from a single blackbody emission
owing to inherent geometric effects of the jet.
For example, the observed temperature is latitude-dependent
due to the Doppler shift and latitude-dependent photospheric radius (\citealp{Peer_A-2008-ApJ.682.463P}).
Therefore, the photospheric emission is better represented
by the emission from a multicolor blackbody (mBB) instead of a single blackbody.
\cite{Ryde_F-2010-Axelsson_M-ApJ.709L.172R} suggest that the observed spectral flux at photon energy $E$
from the photospheric emission can be
described with (i.e., mBB)
\begin{equation}\label{Eq:mBB}
F^{\rm mBB}(E, T_{\rm max }) = \int_{T_{\rm{min}}}^{T_{\rm max }}  \frac{dA(T)}{dT} \frac{E^3}{\exp(E/kT)-1}dT,
\end{equation}
where $T_{\rm{max}}$ is a free parameter, and $T_{\rm{min}} \ll T_{\rm{max}}$ cannot be determined.
For each Planck function,
the spectrally integrated flux for each Planck function is given by $f(T)=A(T)T^4 \pi^4/15$, where $A(T)$ is the normalization as a free parameter.
In \cite{Ryde_F-2010-Axelsson_M-ApJ.709L.172R},
the contribution of each single blackbody emission is took as
\begin{equation}\label{eq:T_PL}
f(T)=f_{\rm{max}}(T/T_{\rm{max}})^q,
\end{equation}
where $f_{\rm{max}}$ is the spectrally integrated flux at $T=T_{\rm{max}}$ and the index $q$ is a free parameter.

In reality, an individual photon in the jet can be scattered to
an observer by an electron at any position in the outflow with a certain probability.
That is to say, the observed photons can be from both the different latitude
and different radius of the jet.
Since the observed probability $P$ of a photon is related to the optical depth $\tau$ for photons from their location to an observer, i.e., $P\propto \exp(-\tau)$,
a power-law distribution of temperature for mBB may be not well describe the contribution of the emission from different radius of the jet.
By fitting the radiation spectrum of the photosphere from numerical calculation (e.g., \citealp{Deng_W-2014-Zhang_B-ApJ.785.112D}; \citealp{Wang_K-2020-Lin_DB-ApJ.899.111W}),
we point out the temperature distribution of the mBB used to well describe the photospheric emission may be
\begin{equation}\label{eq:T_CPL}
f(T)=f_{\rm max} (T/T_{\rm{c}})^q \exp[-(T/T_{\rm{c}})^s],
\end{equation}
where $T_{\rm c}$ is a cutoff temperature.
The mBB with temperature distribution as Equation~(\ref{eq:T_CPL}) or (\ref{eq:T_PL})
is denoted as CPL-mBB or PL-mBB, respectively.
In the left panel of Figure~{\MyFigA},
one can find that the CPL-mBB with $q=3$ and $s=1.2$ can well model the emission of a non-dissipated photosphere with $r_{\rm ph }>r_{\rm s}$.
Here, $T_{\rm max}\gg T_{\rm c}$ is take in CPL-mBB,
$r_{\rm ph }$ is the radius of the photosphere, and $r_{\rm s}$ is the saturation radius of the fireball.
In the right panel of Figure~{\MyFigA}, we shows the time-resolved $EF_E$ spectrum fitting with CPL-mBB and power-law spectral model for one of the time intervals in table~1 of \cite{Ryde_F-2010-Axelsson_M-ApJ.709L.172R}, i.e., [9.22, 9.47]~s. the value of C-stat/dof=286.75/286 is reported in the fitting.

We would apply the CPL-mBB with $q=3$ and $s=1.2$ to fit the thermal component in other GRBs.
Based on the fitting result with CPL-mBB, one can obtain the value of $T_{\rm c}$ and $f_{\rm max}$.
The photospheric temperature $T_{\rm ph}$ of the outflow propagating along the light of sight
is related to the value of $T_{\rm c}$ by $T_{\rm ph} = 3 T_{\rm c}$ according the numerical results.
Correspondingly, $F_{\rm ph}\equiv\sigma T_{{\mathop{\rm ph}\nolimits} }^4r_{{\rm{ph}}}^2/d_L^2= 0.63\int_{0}^{+\infty}{F^{\rm mBB}dE}$ according the numerical results.
The values of $T_{\rm ph}$ and $F_{\rm ph}$ can be used to estimate the properties of the photosphere.

\section{Data Reduction and Spectral Analysis of GRB~220426A}\label{Sec:Analysis}
GRB~220426A was detected by \emph{Fermi} Gamma-Ray Burst Monitor (GBM) at 07:14:08 UT ($T_0$)
on 2022 April 26 with duration $T_{90}\sim 6$~s estimated in energy band of 50-300~keV (\citealp{Malacaria_C-2022-Meegan_C-GCN.31955.1M}).
GBM has 12 sodium iodide (NaI) scintillation detectors covering the 8~keV-1~MeV energy band,
and two bismuth germanate (BGO) scintillation detectors that are sensitive to the 200~keV-40~MeV energy band
(\citealp{Meegan_C-2009-Lichti_G-ApJ.702.791M}). The brightest NaI and BGO detectors, i.e., NaI2 and BGO0, are used for our analyses.
The light-curve of GRB~220426A can be found in Figure~{\MyFigB}.
From these light-curves, one can find that the burst consists of a double main pulses with some sub-peaks.

As reported by \cite{Biltzinger_B-2022-Kunzweiler_F-GCN.31950.1B},
the time-averaged spectrum from $0.002$~s to $9.856$~s can be fitted with Band function
and the best fit result of Band function is $E_{\rm p}=146.3\pm0.9$~keV, $\alpha=-0.05\pm0.01$,
and $\beta=-3.08\pm 0.04$.
It reveals that the Band component in this burst is very narrow.
Then, we would like perform the detail spectral analysis of the prompt emission in this burst.
Firstly, the Band function is used in our spectral fitting.
The fitting results can be found in Figure~{\MyFigC} and reported in 3-7 columns of Table~{\MyTabA},
where different time intervals, i.e., [0, 6]~s, [0, 1]~s, [1, 2]~s, [2, 3]~s, [3, 4]~s, [4, 5]~s, and [5, 6]~s,
are adopted.
Here, the time intervals of [0, 1]~s, [1, 2]~s, [2, 3]~s, [3, 4]~s, [4, 5]~s, and [5, 6]~s are indicated in Figure~{\MyFigB} with labels of a, b, c, d, e, and f, respectively.
In the time interval [0, 6]~s, the best fitting result with Band function reports $E_0=76$~keV, $\alpha=0.1$,
and $\beta=-3.28$, which is consistent with the result reported in \cite{Biltzinger_B-2022-Kunzweiler_F-GCN.31950.1B}.
In the other time intervals, the spectral fittings also reveal a narrow Band component with high value of $\alpha\sim 0.3$ and low value of $\beta\sim -3.4$.
Both high $\alpha$ and low $\beta$ are strong reminiscent of GRB~090902B,
of which the spectral analysis with Band function
reported $\alpha\gtrsim -0.3$ and $\beta\lesssim -3.7$
in the time intervals of [0, 13]~s after the \emph{Fermi} trigger.
The narrow Band component of GRB~090902B is found to be consistent with a
multi-color quasi-thermal spectrum (\citealp{Ryde_F-2010-Axelsson_M-ApJ.709L.172R}; \citealp{Zhang_BB-2011-Zhang_B-ApJ.730.141Z}).
Then, we would like believe that the narrow Band component found in GRB~220426A may be also consistent
with a multi-color quasi-thermal spectrum.
In Figure~{\MyFigD}, we show the spectral fitting results of GRB~220426A with CPL-mBB, $q=3$, and $s=1.2$.
The fitting results are also reported in 8-10~columns of Table~{\MyTabA}.
According to Table~{\MyTabA},
one can find that the value of C-stat/dof in the fitting with CPL-mBB
is closer to unit than that in the fitting with Band function.
This indicate that the CPL-mBB with $q=3$ and $s=1.2$
can present a better description of the radiation spectrum observed in GRB~220426A compared with Band function.
It reveals that the prompt emission of GRB~220426A originates from the photosphere.

\section{Conclusion and Discussion}\label{Sec:Conclusion}
In this paper, we report on a very bright, long-duration gamma-ray burst, GRB~220426A,
observed by \emph{Fermi} satellite.
With total duration of $T_{90}=6$~s,
GRB~220426A is composed with two main pulses and some sub-peaks.
The spectral analysis of this burst reveals very narrow Band component around 100~keV,
which is strong reminiscent of GRB~090902B.
Then, we perform the detail spectral analysis of this burst
based on model of a non-dissipated photospheric emission.
Here, the emission of a non-dissipated photospheric emission
is modelled as the multicolor-blackbody with a cut-off power-law distribution of the thermal temperature.
It is found that the photospheric emission can present a better description about
the radiation spectrum of this burst compared with Band function.
Then, we conclude that this burst would be a second burst
in the class of GRB~090902B observed by \emph{Fermi} satellite.

The identification of the emission from the photosphere
allows one to determine physical properties of the relativistic
outflow, such as the bulk Lorentz factor $\Gamma_{\rm ph}$ and the radius $r_{\rm ph}$ of the photosphere,
and the initial size of the flow $r_0$ (\citealp{Peer_A-2007-Ryde_F-ApJ.664L.1P}).
In the case of $r_{\rm ph}>r_{\rm s}=\Gamma r_0$,
the photospheric radius is given by $r_{\rm ph}= P_{\rm jet} \sigma_{\rm T} / 8 \pi \Gamma^3 m_{\rm{p}} c^3$,
where $P_{\rm jet}$ is the power of the jet.
In addition, one can have $P_{\rm jet} = 4 \pi d_{\rm L}^2 (r_{\rm ph}/r_{\rm s})^{2/3}F_{\rm ph}$,
${T_{{\rm{ph}}}} = {T_0}{\left( {{r_{{\rm{ph}}}}/{r_{\rm{s}}}} \right)^{ - 2/3}}$,
and ${T_0} = {\left[ {{P_{{\rm{jet }}}}(t)/(16\pi r_0^2\sigma )} \right]^{1/4}}$.
Taking $Y$ to represent the radiation efficiency of the jet in the gamma-rays,
one can have $P_{\rm jet} = 4\pi d_{\rm L}^2Y F_{\rm obs}$, where $F_{\rm obs}$ is the observed spectrally integrated gamma-ray flux
and can be estimated with $F_{\rm obs}\simeq F_{\rm ph}$ in GRB~220426A.
Then, one can use $kT_{\rm ph}$ and $F_{\rm ph}$ to estimate $r_{\rm ph}$ and $\Gamma_{\rm ph}$.
In Figure~{\MyFigE}, we shows the evolution of $kT_{\rm ph}$, $F_{\rm ph}$, $r_{\rm ph}$, and $\Gamma_{\rm ph}$
in the course of the burst. Since the redshift $z$ of GRB~220426A is not known, we take $z=1$.
One can find that the temperature and Lorentz factor of the photosphere decrease with time.
However, the radius of the photosphere increases with time.

We also note that there seems to be an additional component in the radiation spectrum.
Then, we would like to fit the radiation spectrum with CPL-mBB and power-law spectral model $F(E)=F_0E^{-\beta}$.
The fitting results is shown in the left panel of Figure~{\MyFigF},
where the time interval of [1, 2]~s is fitted as an example.
The best fitting result reveals that the power-law spectral component is presented with $\beta\sim -2$.
The power-law spectral model with $\beta\sim -2$ is very similar to the spectral shape in the low-energy regime of the radiation spectrum in CPL-mBB model.
It may reveal that the high-energy emission in this burst may be formed by
inverse Compton of the CPLmBB photons by the electrons.
Then, we perform the calculation of the inverse Compton of the CPL-mBB photons by the electrons with different $\gamma_e$ and ${{\sigma _{\rm{T}}}{n_e}}=0.01$, where $\sigma _{\rm{T}}$ is the Thomson cross-section and $n_e$ is the number of electrons along the line of sight.
The inverse Compton component can be found in the right panel of Figure~{\MyFigF}.
One can find that the spectral shape of the inverse Compton component
in the high-energy regime can be described with a power-law function with $F(E)\propto E^{\sim 2}$.

\acknowledgments
This work is supported by
the Guangxi Science Foundation (grant Nos. 2018GXNSFFA281010, 2018GXNSFGA281007, 2018GXNSFDA281033)
and the National Natural Science Foundation of China (grant Nos.11773007, 12133003, 11503011, U1938201, U1938106).

\clearpage
\begin{table}
\caption{Spectral fittings of GRB~220426A with Band function (3-7columns) and CPLmBB with $q=3$ and $s=1.2$ (8-10 columns)}\label{MyTabA}
    \centering
    \resizebox{\textwidth}{12mm}{
    \begin{tabular}{cc|ccccc|ccc}
    \hline \hline
    Time Interval& Labels & $\alpha$ & $\beta$ & $E_0$(keV) & norm & C-stat/dof & $kT_{c}$(keV) & $f_{\rm max}$ & C-stat/dof \\
    \hline
    $[0,6.0]$~s & $ $ & $0.11\pm0.01$ & $-3.25\pm0.05$ & $75.04\pm0.94$ & $1.67\pm0.03$ & $627.97/357$ & $17.39\pm0.06$ & $2338\pm9.56$ & $707.32/359$ \\
    $[0,1.0]$~s   & $a$ & $0.68\pm0.08$ & $-3.44\pm0.29$ & $83.03\pm4.64$ & $0.52\pm0.04$ & $313.00/357$ & $28.78\pm0.53$ & $1283\pm26.83$ & $340.49/359$\\
    $[1.0,2.0]$~s & $b$ & $0.41\pm0.03$ & $-3.71\pm0.21$ & $77.05\pm2.05$ & $1.60\pm0.07$ & $363.50/357$ & $22.03\pm0.20$ & $2686\pm28.76$ & $358.22/359$\\
    $[2.0,3.0]$~s & $c$ & $0.21\pm0.03$ & $-3.55\pm0.15$ & $92.00\pm2.12$ & $1.71\pm0.05$ & $398.11/357$ & $22.68\pm0.17$ & $3848\pm34.02$ & $372.51/359$\\
    $[3.0,4.0]$~s & $d$ & $0.20\pm0.03$ & $-4.07\pm0.28$ & $65.88\pm1.62$ & $2.28\pm0.10$ & $342.02/357$ & $16.03\pm0.13$ & $2416\pm22.58$ & $322.64/359$\\
    $[4.0,5.0]$~s & $e$ & $0.19\pm0.03$ & $-3.73\pm0.13$ & $57.97\pm1.21$ & $4.00\pm0.16$ & $408.71/357$ & $14.13\pm0.09$ & $3231\pm23.90$ & $369.14/359$\\
    $[5.0,6.0]$~s & $f$ & $-0.09\pm0.04$ & $-4.31\pm0.50$ & $50.02\pm1.77$ & $1.86\pm0.14$ & $275.67/357$ & $9.89\pm0.10$  & $1107\pm12.36$ & $271.55/359$\\
    \hline
    \end{tabular}}
\end{table}

\clearpage
\begin{figure}
\centering
\begin{tabular}{cc}
\centering
\includegraphics[angle=0,scale=0.30]{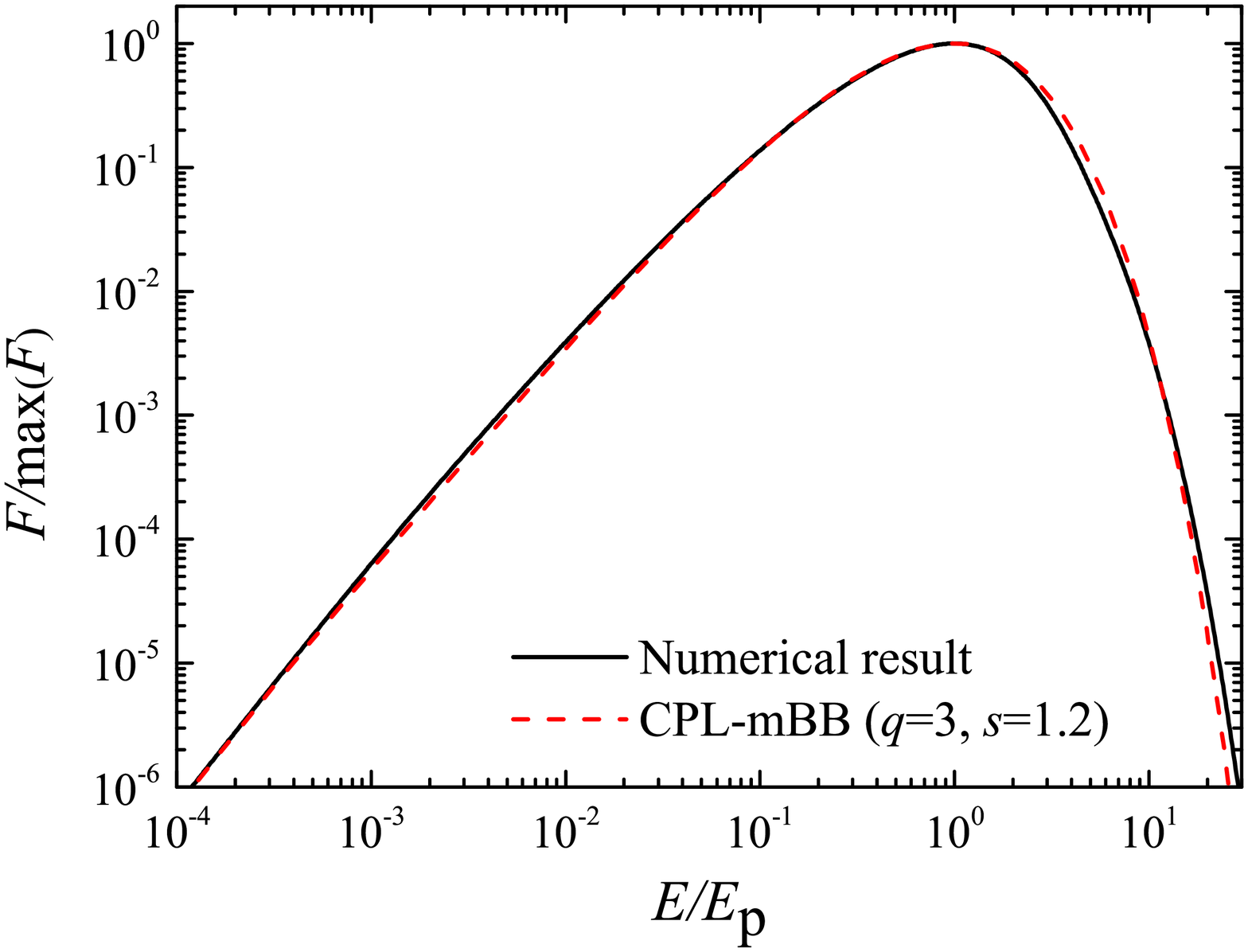} &
\includegraphics[angle=0,scale=0.79, trim=0 25 0 10]{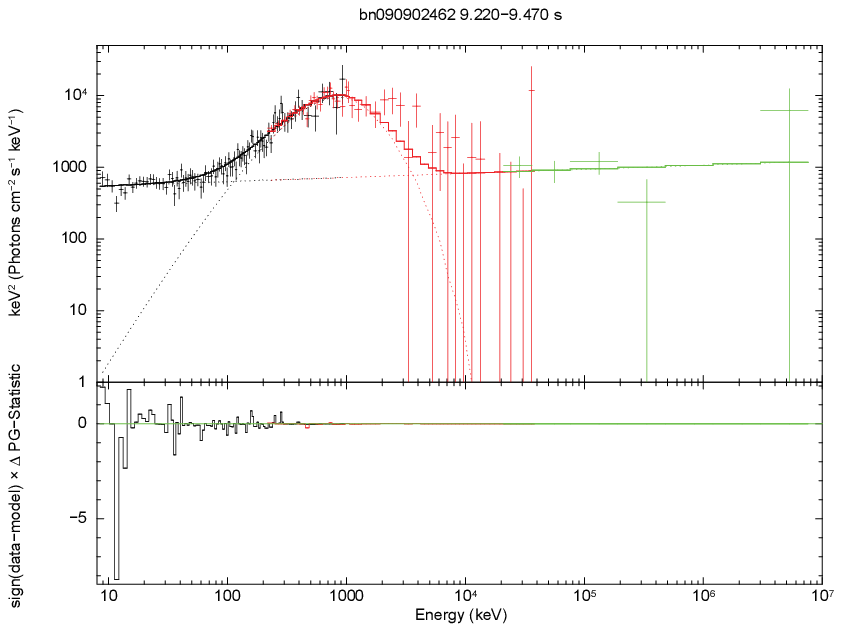}\\
\end{tabular}
\caption{\emph{Left panel}: Modeling the radiation spectrum of photospheric emission (numerical result, solid line)
with CPL-mBB (dashed line); \emph{Left panel}: Fitting the radiation spectrum of GRB~090902B with CPL-mBB and power-law spectral model.}\label{MyFigA}
\end{figure}

\begin{figure}
\centering
\begin{tabular}{cc}
\centering
\includegraphics[angle=0,scale=0.50]{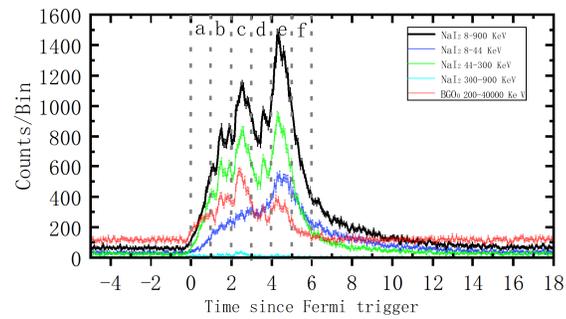}
\end{tabular}
\caption{Light-curves of GRB~220426A in different energy ranges.}\label{MyFigB}
\end{figure}

\clearpage
\begin{figure}
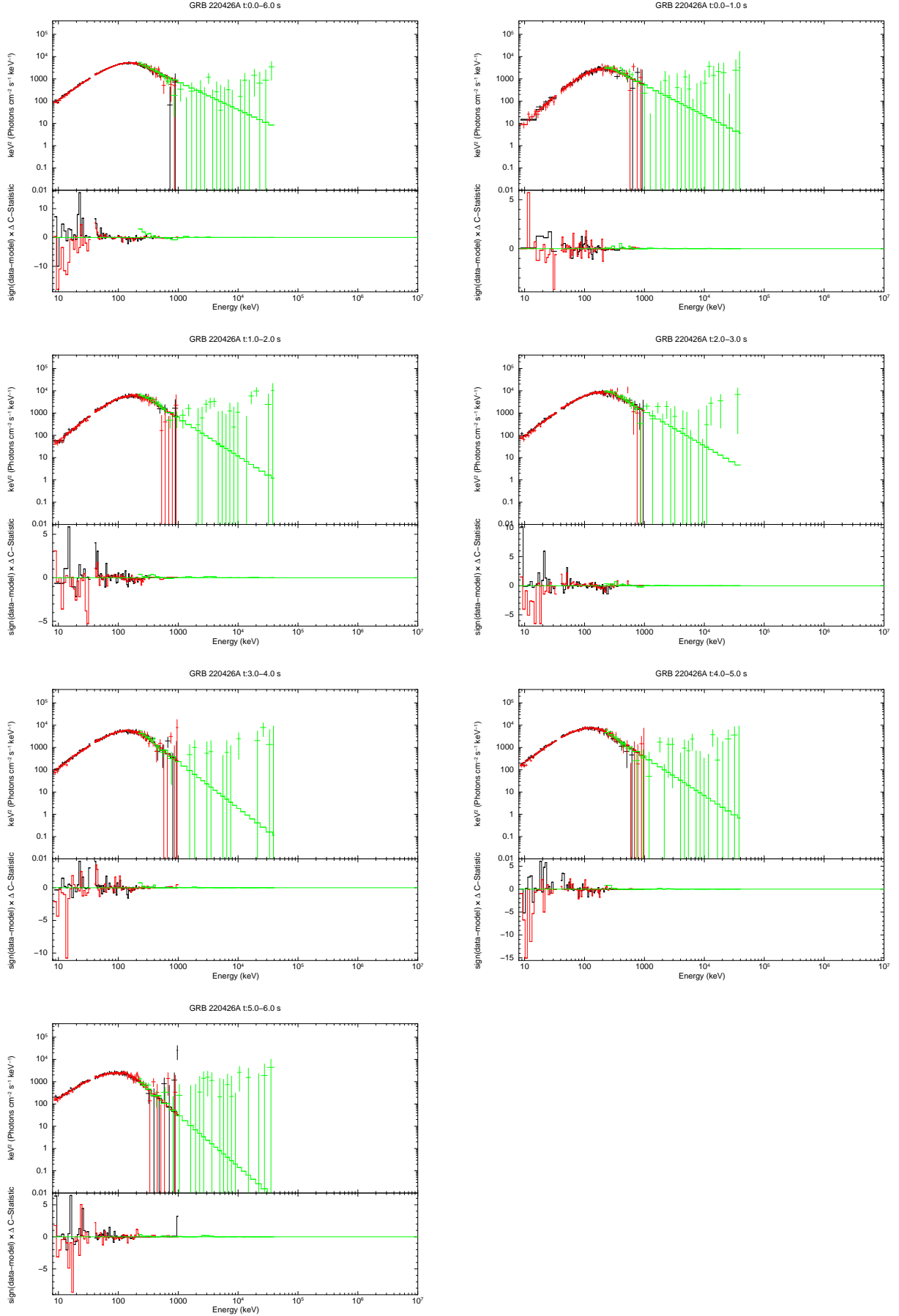

\centering
\begin{tabular}{cc}
\includegraphics[angle=-90,width=0.45\textwidth]{Fig_band0-6.eps} &
\includegraphics[angle=-90,width=0.45\textwidth]{Fig_band0-1.eps} \\
\includegraphics[angle=-90,width=0.45\textwidth]{Fig_band1-2.eps} &
\includegraphics[angle=-90,width=0.45\textwidth]{Fig_band2-3.eps} \\
\includegraphics[angle=-90,width=0.45\textwidth]{Fig_band3-4.eps} &
\includegraphics[angle=-90,width=0.45\textwidth]{Fig_band4-5.eps} \\
\includegraphics[angle=-90,width=0.45\textwidth]{Fig_band5-6.eps}  &
\\
\end{tabular}
\caption{Spectrum analysis of GRB 220426A for the time intervals of [0,6]~s, [0, 1]~s, [1, 2]~s, [2, 3]~s, [3,4]~s, [4, 5]~s, and [5, 6]~s, respectively. Here, a band function is used in our fittings.}
\label{MyFigC}
\end{figure}

\clearpage
\begin{figure}
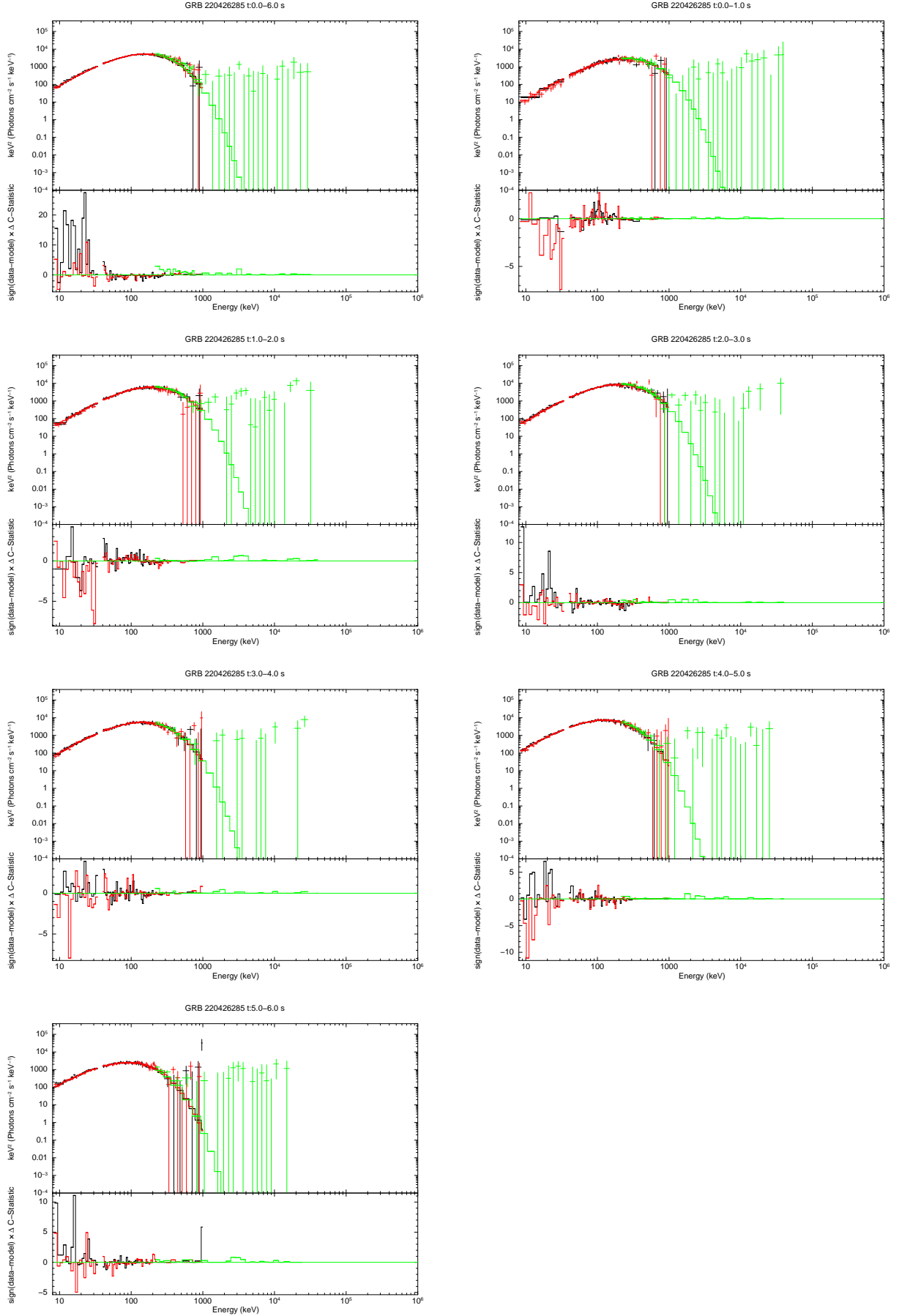

\centering
\begin{tabular}{cc}
\includegraphics[angle=-90,width=0.45\textwidth]{Fig_cplmbb0-6}&
\includegraphics[angle=-90,width=0.45\textwidth]{Fig_cplmbb0-1}\\
\includegraphics[angle=-90,width=0.45\textwidth]{Fig_cplmbb1-2}&
\includegraphics[angle=-90,width=0.45\textwidth]{Fig_cplmbb2-3}\\
\includegraphics[angle=-90,width=0.45\textwidth]{Fig_cplmbb3-4}&
\includegraphics[angle=-90,width=0.45\textwidth]{Fig_cplmbb4-5}\\
\includegraphics[angle=-90,width=0.45\textwidth]{Fig_cplmbb5-6}&
\\
\end{tabular}
\caption{Same as Figure~\ref{MyFigC}, but with CPL-mBB spectral model in the fittings. }
\label{MyFigD}
\end{figure}

\clearpage
\begin{figure}
\centering
\begin{tabular}{cc}
\includegraphics[width=0.45\textwidth]{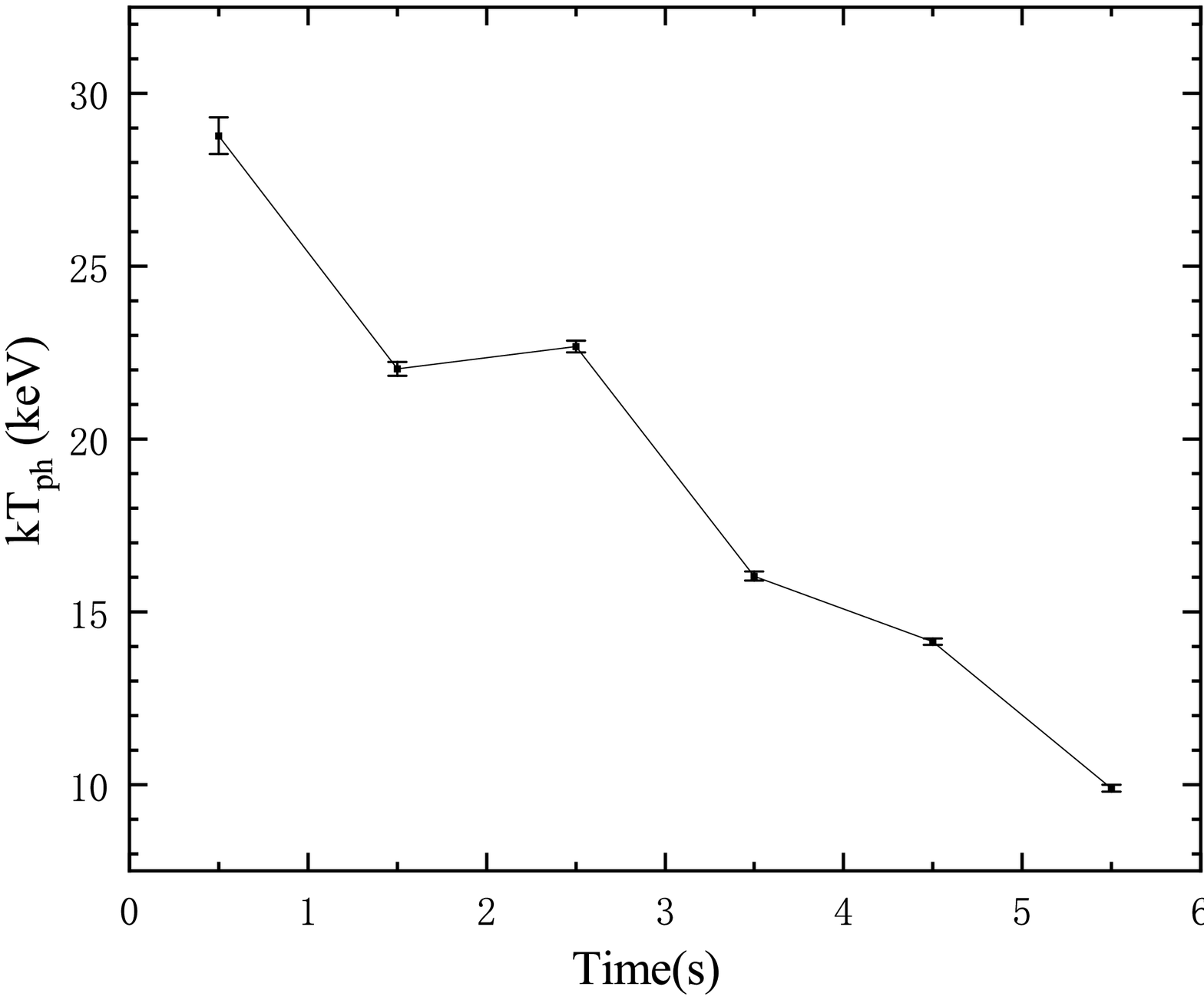}&
\includegraphics[width=0.45\textwidth]{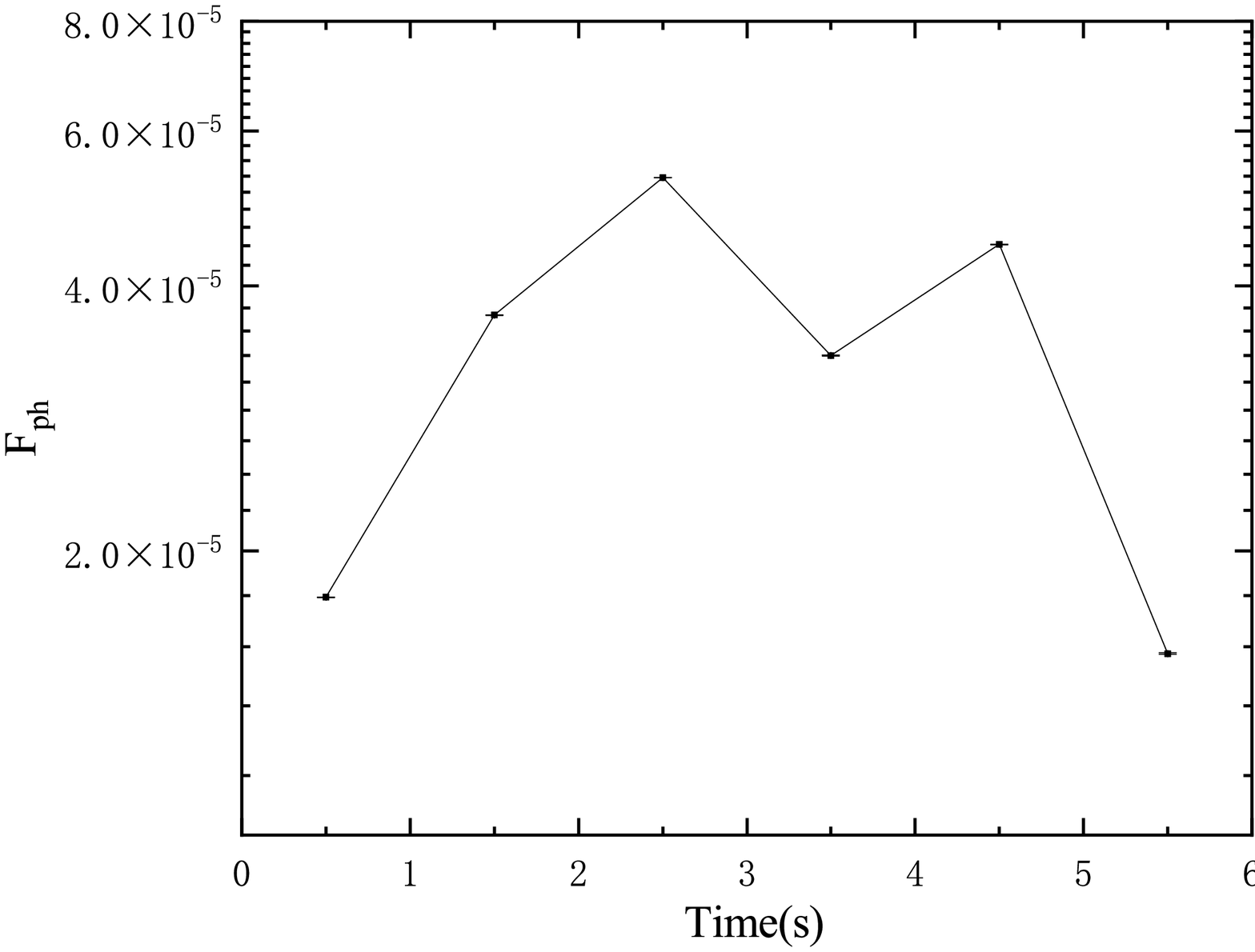}\\
\includegraphics[width=0.45\textwidth]{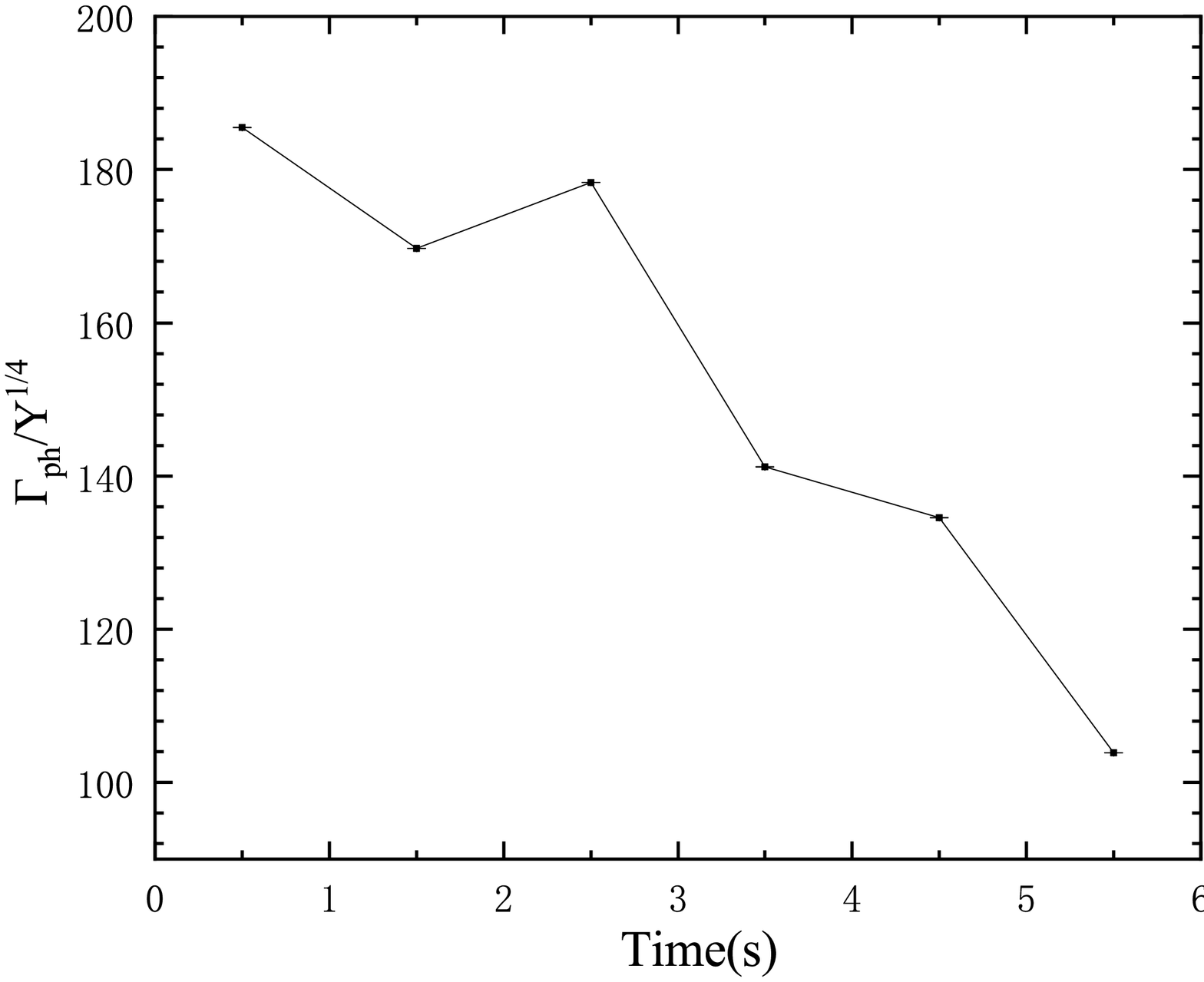}&
\includegraphics[width=0.45\textwidth]{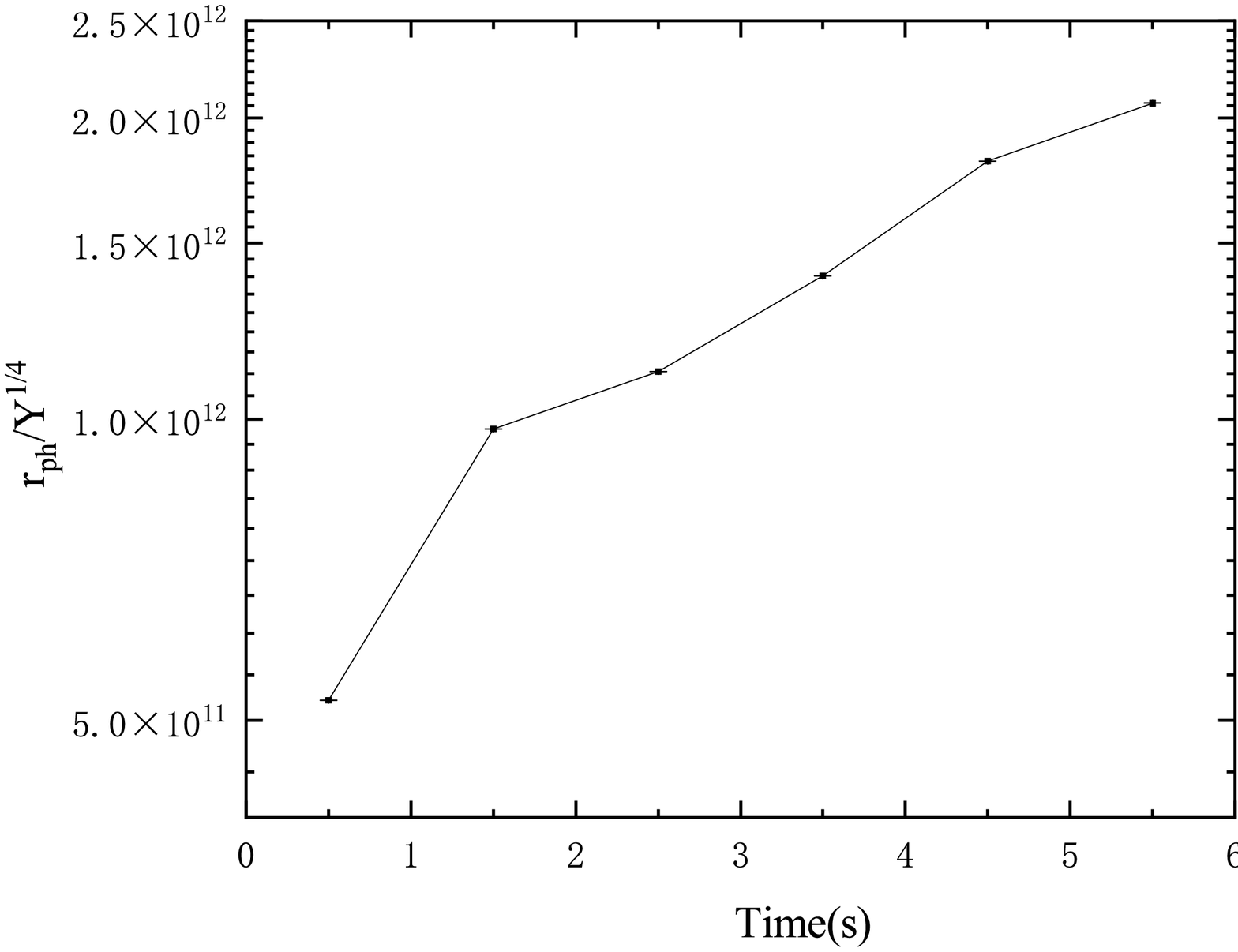}\\
\end{tabular}
\caption{Evolution of $kT_{\rm ph}$, $F_{\rm ph}$, $r_{\rm ph}$, and $\Gamma_{\rm ph}$
in the course of the burst.}
\label{MyFigE}
\end{figure}

\clearpage
\begin{figure}
\centering
\begin{tabular}{cc}
\includegraphics[scale=0.95, trim=0 27 0 8]{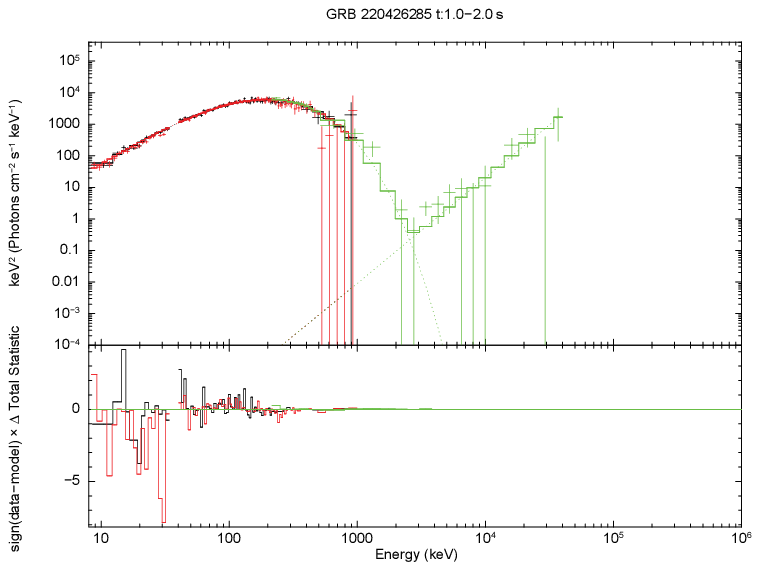} &
\includegraphics[scale=0.33]{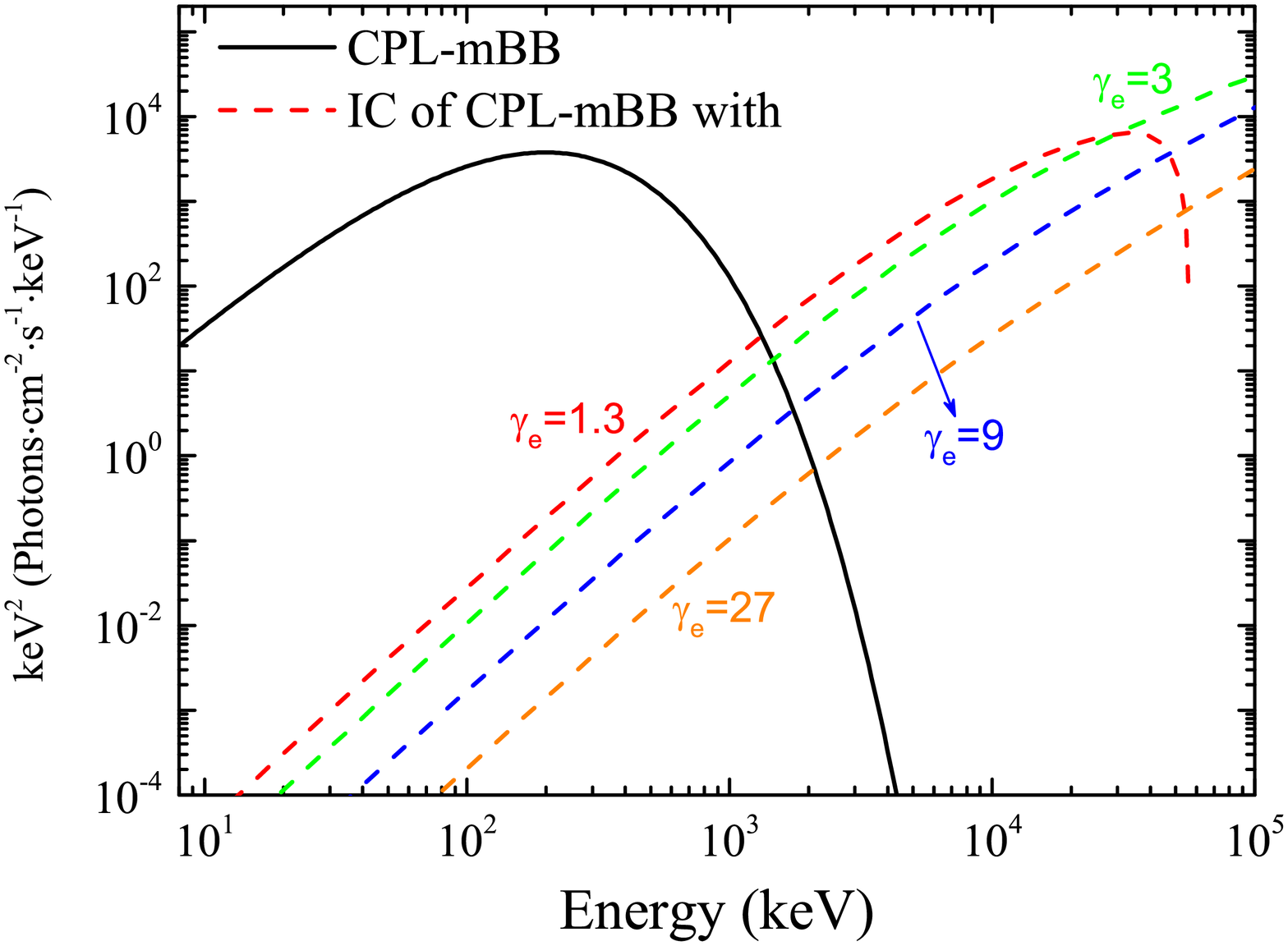} \\
\end{tabular}
\caption{Spectral fitting results with CPLmBB and power-law spectral model (left panel) and theoretical inverse Compton spectrum (right panel).}
\label{MyFigF}
\end{figure}

\end{document}